 \documentclass[aps,prl,twocolumn,superscriptaddress,showpacs,amsmath,amssymb]{revtex4}

\usepackage{graphicx}
\usepackage{dcolumn} 
\usepackage{bm}      
\usepackage{verbatim}%
\usepackage{epstopdf}

\begin{document}


\title
{
Molecular Dynamics Simulation of Ga Penetration along Grain Boundaries in Al: a Dislocation Climb Mechanism
}



\author{Ho-Seok Nam}
\email
{hnam@princeton.edu}

\affiliation
{
Department of Mechanical and Aerospace Engineering, Princeton University, Princeton, NJ 08544
}

\author{D. J. Srolovitz}

\affiliation
{
Department of Mechanical and Aerospace Engineering, Princeton University, Princeton, NJ 08544
}
\affiliation
{
Department of Physics, Yeshiva University, New York, NY 10033
}


\date{\today}

\begin{abstract}

Many systems where a liquid metal is in contact with a polycrystalline solid exhibit deep liquid grooves where the grain boundary meets the solid-liquid interface.  For example, liquid Ga quickly penetrates deep into grain boundaries in Al, leading to intergranular fracture under very small stresses.  We report on a series of molecular dynamics simulations of liquid Ga in contact with an Al bicrystal.  We identify the mechanism for liquid metal embrittlement, develop a new model for it, and show that is in excellent agreement with both simulation and experimental data.

\end{abstract}










\maketitle


When liquid metals are brought into contact with other polycrystalline metals, deep liquid-filled grooves often form at the intersections of grain boundaries and the solid-liquid interface.  In some systems, such as Al-Ga, Cu-Bi and Ni-Bi, the liquid film quickly penetrates deep into the solid along the grain boundaries and leads to brittle, intergranular fracture under the influence of modest stresses.  This is a form of liquid metal embrittlement (LME).  This phenomenon is ubiquitous in material processing and is particularly important in nuclear reactor scenarios in which liquid metals are used as coolants and as spallation targets.  

Al-Ga is a particularly well-studied LME system.  Transmission electron microscopy (TEM)~\cite{Hugo:AlGaTEM}, scanning electron microscopy (SEM)~\cite{Kozlova:AlGaSEM}, and synchrotron radiation microtomography~\cite{Pereiro-Lopez:AlGaPRL2005, Pereiro-Lopez:AlGaSRMR} studies all show that liquid Ga penetrates into grain boundaries in Al at a remarkable rate.  Ahead of the liquid Ga-grain boundary groove tip, the grain boundary is wetted by a Ga layer of thickness ranging from several monolayers~\cite{Hugo:AlGaTEM} to several hundred nanometers~\cite{Pereiro-Lopez:AlGaPRL2005}, even in the absence of an applied stress.  Interestingly, the rate of propagation of the liquid Ga layer along the grain boundary is strongly influenced by even very small stresses~\cite{Hugo:AlGaTEM, Kozlova:AlGaSEM, Pereiro-Lopez:AlGaPRL2005, Pereiro-Lopez:AlGaSRMR}. These observations have led to the conclusion that liquid Ga embrittlement of Al is caused by rapid liquid Ga penetration.  

Several models have been proposed to explain the driving forces and mechanisms by which the liquid phase penetrates quickly along grain boundaries, including mixed diffusion-dissolution~\cite{Bokstein:DiffusionDisolution}, dissolution-reprecipitation~\cite{Glickman:DCM}, coherency stresses~\cite{Rabkin:CoherencyStresses}, and others~\cite{Joseph:LMEreview}.  While each of these approaches is capable of explaining one or more aspects of LME, each also leads to discrepancies with respect to other observed LME phenomena in the same materials system.  For example, none of these approaches successfully explains the effects of stress on liquid film penetration.  Further issues include whether LME is essentially ``replacement-like'' (Ga atoms replace Al atoms at the grain boundary and the Al atoms are transported away) or ``invasion-like''~\cite{Pereiro-Lopez:AlGaPRL2005} (Ga atoms insert interstitially into the grain boundary without replacing Al atoms).    

The penetration of a liquid phase along the grain boundary is a complex phenomenon, involving several different types of simultaneous processes; e.g., dissolution/reprecipitation, liquid groove formation, grain boundary diffusion, grain boundary segregation, \ldots~  The rates at, and degrees to, which these processes occur are associated with  material properties as diverse as solubility in the liquid and solid, solid-liquid interface tension, grain boundary energy, heats of segregation, grain boundary diffusivity, etc.  The tendency for and rate of LME are also sensitive to externally controllable factors such as temperature and applied stress.  Because of the interplay between the underlying phenomena that occur in LME, it has been difficult to design and perform experiments that can be easily interpreted to understand which processes control LME and which are simply parasitic.  In this Letter, we study LME by performing molecular dynamics (MD) simulations of an Al bicrystal in contact with liquid Ga (with and without an applied stress) and investigate how Ga penetrates along the grain boundaries during the early stages of the wetting process.  We use the simulation results to propose a new mechanism for LME and compare it with general trends gleaned from a series of LME experimental studies.  

\begin{figure}[!tbp]
\includegraphics[width=0.30\textwidth]{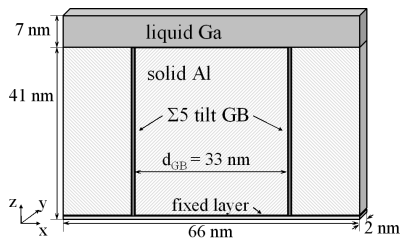}
\caption
{ \label{fig:Geometry}
Simulation cell containing two Al grains ($\sim 320,000$ atoms) in contact with liquid Ga ($\sim 40,000$ atoms).
}
\end{figure}

We describe the atomic interactions using semiempirical embedded-atom method (EAM) potentials for the Al-Ga system that were tuned to successfully reproduce the experimental solid-liquid binary phase diagram~\cite{HoseokNam:LME}.   All of the simulations were performed on a three dimensional Al bicrystal sample in contact with liquid Ga, as shown schematically in  Fig.~\ref{fig:Geometry}.  We impose periodic boundary conditions in the $x$ and $y$ directions, fix several atomic layers at the bottom of the bicrystal (to prevent grain rotation) and leave the top surface free (i.e., there is a vacuum above the liquid).  Periodicity demands that the system contains two identical grain boundaries separated by the grain size $d_{GB}$ in the $x$-direction.  

The MD simulations were performed under constant strain conditions ($NVT$ ensemble), where the strain was chosen to provide the desired stress in the bulk bicrystal~\cite{HoseokNam:LME}.  We perform the simulations at fixed displacement rather than fixed load in order to mimic the effect of a very thick polycrystalline sample. The simulations were performed at $T=600$ K with a uniaxial stress, $0\le \sigma_{xx}\le 500$ MPa.  Simulation time was at least 50 ns.   

Several different grain boundary types were examined in the present simulations: symmetric and asymmetric $\Sigma 5~[001]$ tilt boundaries, symmetric $\Sigma 5~[100]$ twist boundaries, and low angle tilt boundaries.  We found that the Ga penetration behavior is sensitive to grain boundary type and structure: no grain boundary wetting was observed within our 50 ns simulations for the low angle grain boundaries or the symmetric $\Sigma 5~[100]$ twist boundaries, while the $\Sigma 5~ [001]$ tilt boundaries showed remarkable Ga penetration rates with the application of an applied tensile stress.  Since the asymmetric tilt boundary (inclination angle $\alpha = 18.4 \,^{\circ}$ relative to the symmetric inclination) exhibited the largest  Ga penetration rate, we focus on this grain boundary in the remainder of this paper.  This boundary has a relatively high energy and is not particularly special.  

\begin{figure}[!tbp]
\includegraphics[width=0.36\textwidth]{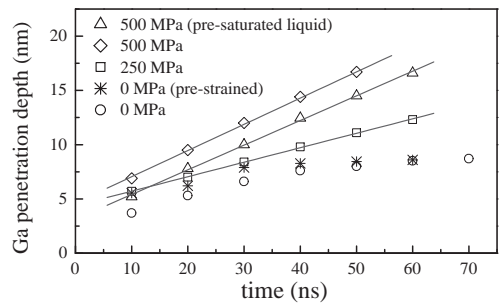}
\caption
{ \label{fig:Prate_asy}
Ga penetration depth versus time.    
}
\end{figure}

We estimate the Ga penetration rate by noting the depth at which the Ga concentration along the grain boundary exceeds a fixed value (one monolayer) at each time~\cite{HoseokNam:LME}.  We plot this depth $L$ versus time $t$ in Fig.~\ref{fig:Prate_asy}.  In the absence of an applied stress, the rate at which Ga penetrates down the grain boundary (slope in Fig.~\ref{fig:Prate_asy}) gradually decreases with time.  However, when stress is applied, the Ga penetration rate becomes nearly time-independent.  Clearly, stress changes the fundamental nature of Ga penetration down grain boundaries in Al: the constant Ga penetration rate suggests that Ga is not simply random walking down the grain boundary ($L \propto t^{1/2}$) nor is the penetration rate controlled by normal grain boundary grooving~\cite{Glickman:DCM} ($L \propto t^{1/3}$ or $t^{1/4}$) but, rather, is strongly driven ($L \propto t$).  The Ga penetration rate increases with applied tensile stress and increasing grain size $d_{GB}$ but is not affected by pre-saturating the liquid with Al.  While stress promotes Ga penetration, it has little effect on the rate of Al dissolution into the liquid~\cite{HoseokNam:LME}.  This implies that dissolution does not control liquid film formation in the Al-Ga system.  

\begin{figure*}[!tbp]
\includegraphics[width=0.11\textwidth]{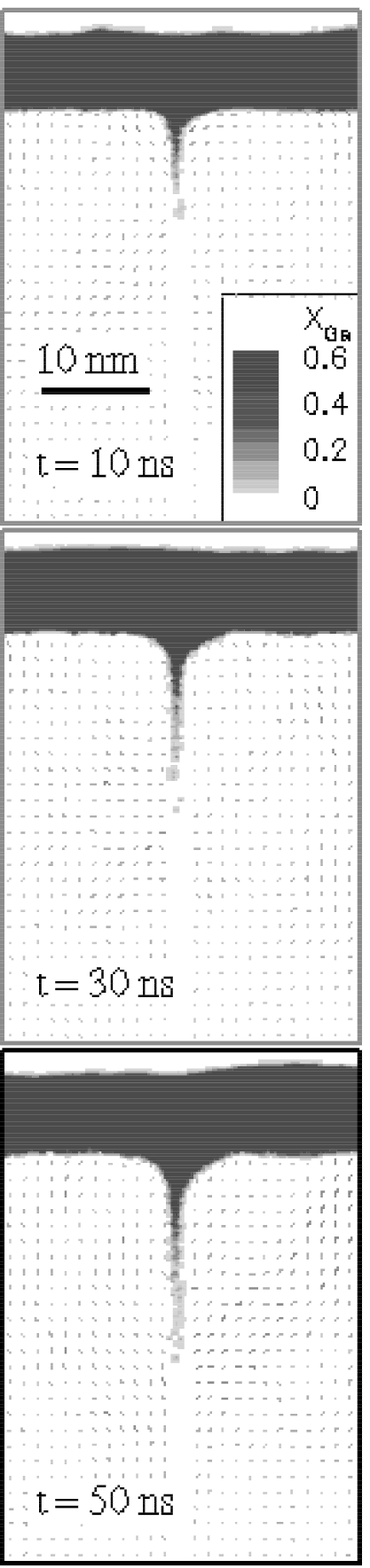}\hspace*{-0.15cm}
\includegraphics[width=0.11\textwidth]{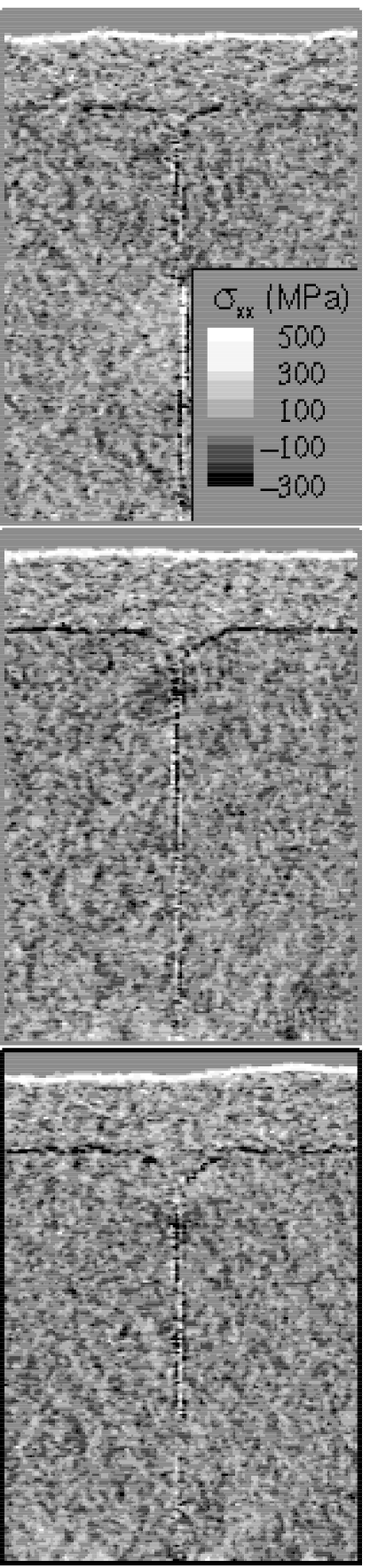}\hspace*{-0.05cm}
\includegraphics[width=0.11\textwidth]{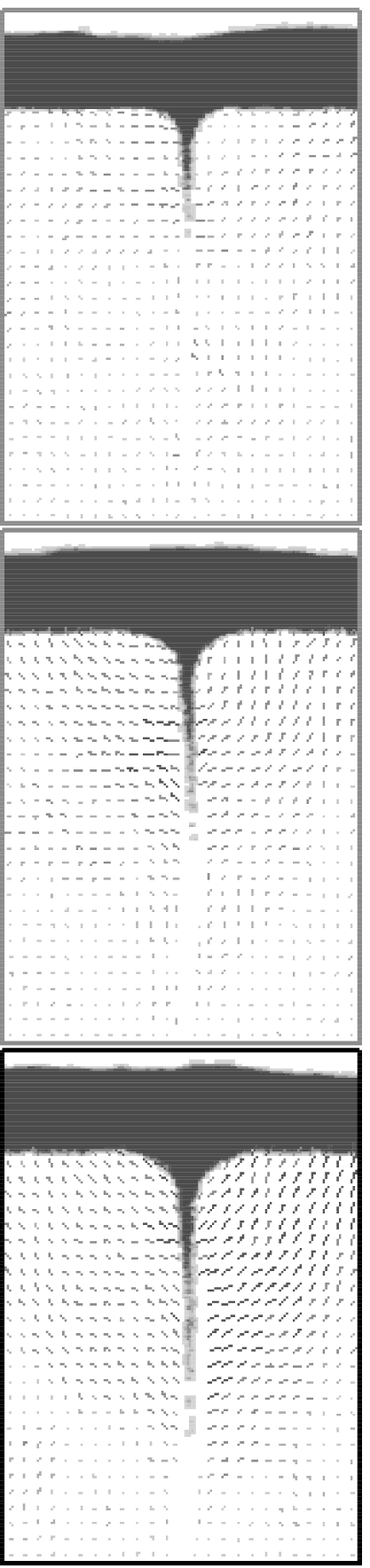}\hspace*{-0.15cm}
\includegraphics[width=0.11\textwidth]{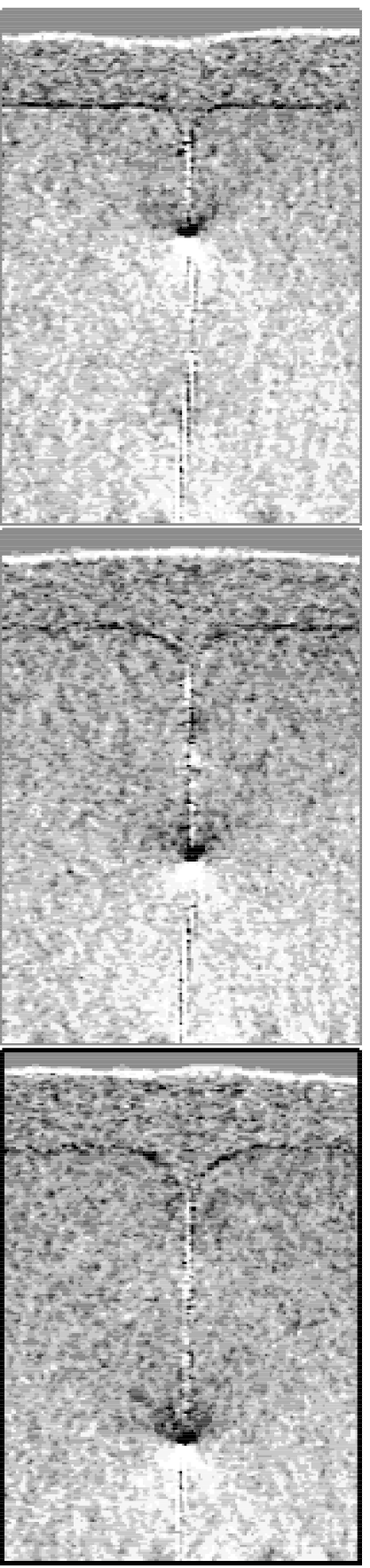}\hspace*{-0.05cm}
\includegraphics[width=0.11\textwidth]{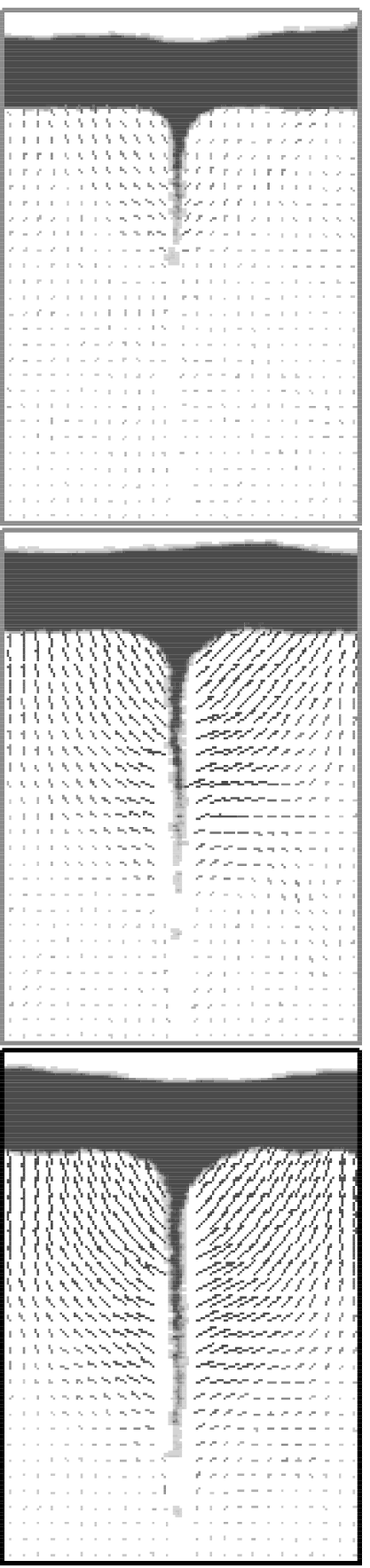}\hspace*{-0.15cm}
\includegraphics[width=0.11\textwidth]{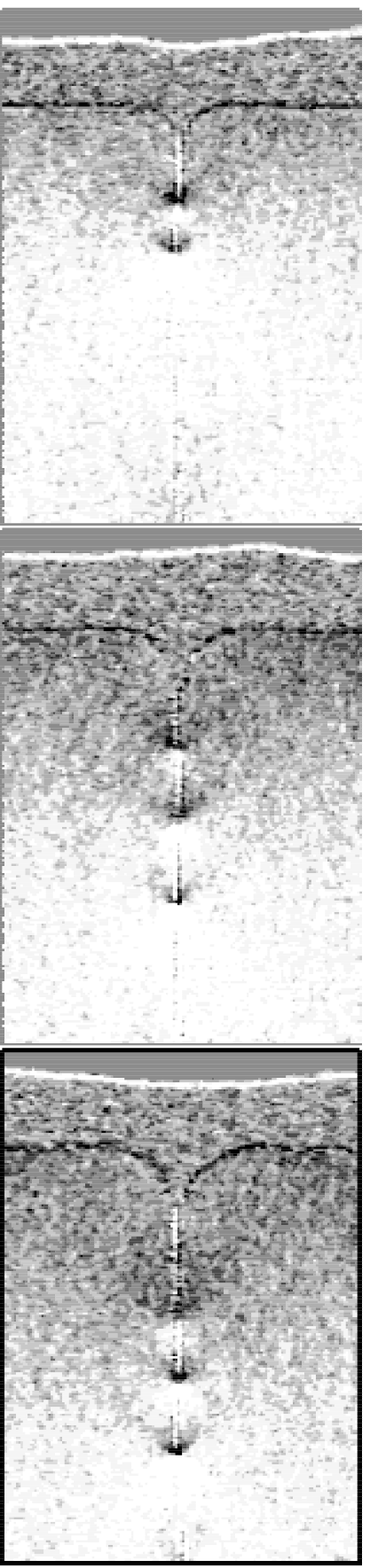}\hspace*{-0.15cm}
\includegraphics[width=0.11\textwidth]{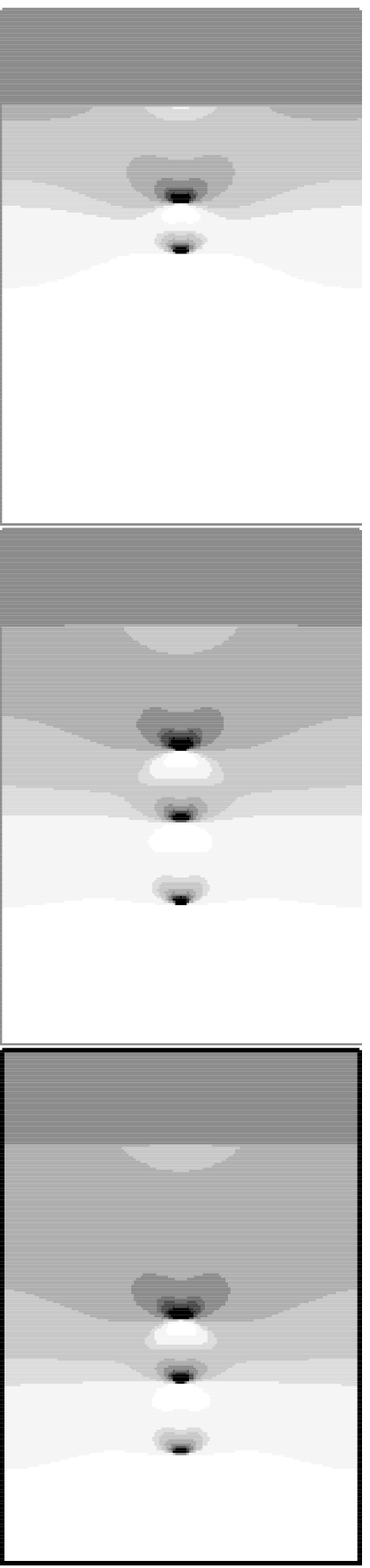}\hspace*{-0.05cm}
\includegraphics[width=0.11\textwidth]{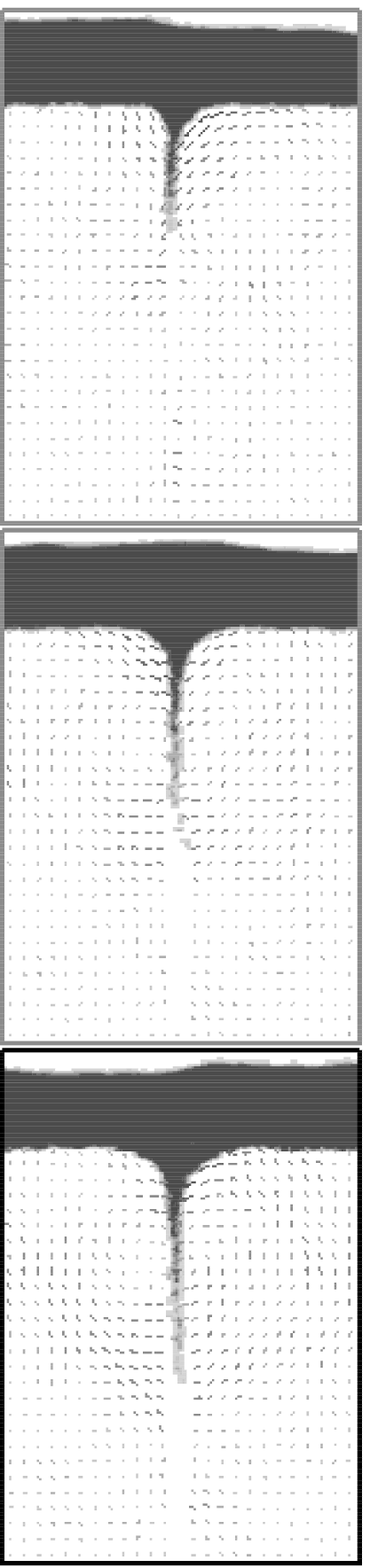}\hspace*{-0.15cm}
\includegraphics[width=0.11\textwidth]{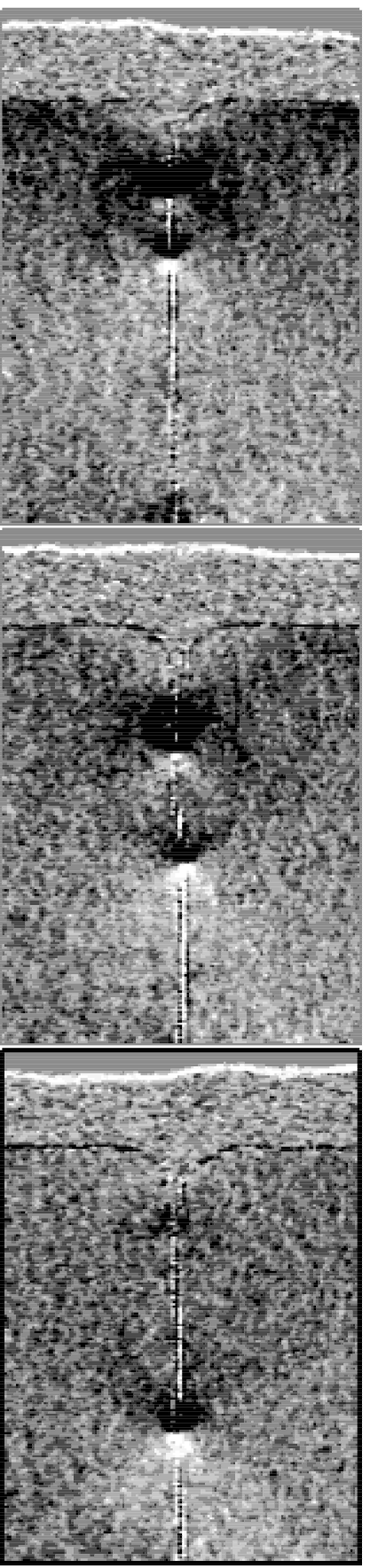}
\newline
\hspace*{0.7cm} (a) 0 MPa \hspace*{2.1cm} (b) 250 MPa \hspace*{3.0cm} (c) 500 MPa \hspace*{2.1cm} (d) 0 MPa (prestrained) \hspace*{0.0cm}
\caption
{ \label{fig:LME_asy}
Liquid metal penetration profiles at $t=$ 10, 30, and 50 ns (from top to bottom) for simulations performed at $T=600$ K.  The left panels of (a)-(d) show the Ga concentration profile (mole fraction $X_{Ga}$) and displacements (vectors - $10\,\times\,$the actual lengths) and the right panels of (a)-(d) show the stress distribution ($\sigma_{xx}$ in MPa) for simulations with an applied stress (a) of 0, (b) 250 MPa, (c) 500 MPa, and (d) of 0 albeit pre-stressed to 500 MPa, respectively.  In (c), stress fields from the dislocation model are added to the right of stress fields from the simulation.  In order to limit consideration to displacements associated with elastic deformation of the solid, we excluded atoms for which the displacements exceeded 5 {\AA} (i.e., primarily diffusive hops) and, as a result, no arrows are plotted in the region of the grain boundary.    
}
\end{figure*}

The atomistic mechanisms operating at the tip of the advancing Ga layers can be identified by analyzing the displacement and stress fields within the system.  Figure~\ref{fig:LME_asy}(a) shows the Ga concentration profile (left) and stress distribution (right) at $t=$ 10, 30, and 50 ns in the absence of an applied stress.  At the beginning of the simulation ($t<10$ ns), some Al atoms at the grain boundaries selectively dissolve into the liquid Ga and a liquid groove forms at the intersection of the grain boundary and the solid-liquid interface.  As the liquid approaches saturation, the liquid groove shape evolution slows and the Ga penetration rate decreases in the absence of an applied stress (see Fig.~\ref{fig:Prate_asy}).  Below the liquid groove root, the Ga concentration decays slowly down the grain boundary rather than abruptly terminating there.   The arrows in the Ga profile plot show the displacement field in the solid, measured with respect to the atom positions at $t=$ 5 ns (the arrow sizes were magnified by a factor of 10).  In the absence of an applied stress, the observed displacements in the solid are very small and random.    

Figures~\ref{fig:LME_asy}(b) and (c) show Ga penetration, displacements in the solid, and stress distribution at constant strains of  $0.65 \%$ ($\sim 250$ MPa) and $1.3 \%$ ($\sim 500$ MPa), respectively.  Although the liquid groove shapes and wetting angle are nearly the same in Figs.~\ref{fig:LME_asy}(a-c), the Ga penetration  is strongly enhanced by the application of stress, forming nanometer-thick Ga-rich films.  In this case, the atomic displacements are considerably larger than in the absence of an applied stress and show the presence of an ordered displacement field (i.e., not just thermal vibrations).  Where Ga has thoroughly penetrated the grain boundary, the displacements are away from the grain boundary.  Additional simulations (not shown) demonstrate that doubling the grain size doubles the displacements and leads to an increase in the grain boundary opening rate.  This, in turn, leads to a dramatic increase in the quantity of Ga at the grain boundary~\cite{HoseokNam:LME}.  Doubling the grain size doubles the strain energy stored in the bicrystal under the present fixed-grip loading.  

Examination of the displacement field confirms that  an applied strain leads to the formation of a thicker penetrating Ga-rich layer.  At the same time, the formation of the nanometer-thick Ga-rich layer helps relax the residual stress.  We can examine the interplay of these by considering the stresses within the system;  $\sigma_{xx}$ is shown to the right of the Ga penetration figures in Fig.~\ref{fig:LME_asy}.  Figure~\ref{fig:LME_asy}(a) shows that in the absence of an applied strain, the stresses in the system are small and random.  However, when a strain is applied, we observe the formation of one [Fig.~\ref{fig:LME_asy}(b)] or more [Fig.~\ref{fig:LME_asy}(c)] patterns of concentrated stress at the grain boundary.  These patterns consist of a dark (large compressive) region above a light (large tensile) region.  This suggests that these stress concentrations are associated with edge dislocations with a Burgers vector perpendicular to the boundary plane.  We can confirm the identification of these stress concentrations by performing an analytical calculation of the stress field associated with such dislocations~\cite{Head:EdgeDislStress}.  A comparison of the simulation and linear elastic stress fields is shown in Fig.~\ref{fig:LME_asy}(c).  By comparing the stress fields, we determine that the magnitude of the Burgers vector to be $\sim 0.5$ \AA;  much smaller than that of a lattice dislocation. 

Figure~\ref{fig:LME_asy}(b) shows that the dislocation, once formed, ``climbs'' down along the grain boundary at a nearly constant rate.  Further examination of Fig.~\ref{fig:LME_asy}(c) (larger strain) shows that the first dislocation climbs down the grain boundary at the same constant rate as the single dislocation in Fig.~\ref{fig:LME_asy}(b) (low strain).  However, in this case, once the first dislocation has moved some distance from its point of origin, second and third dislocations are nucleated one after another, and climb down the grain boundary too, leading to three equally spaced dislocations that all move at the same rate.  Therefore, these special grain boundary dislocations only form above a critical applied strain and ``climb'' down the grain boundary at a constant rate that is independent of the magnitude of this strain.  Increasing applied strain simply results in the formation of more dislocations.  Increasing the grain size also results in the formation of more dislocations at the same applied strain.  

Why do the dislocations move down?  The dislocation sets up its own stress field; in the present geometry, it is compressive above the dislocation and tensile below.  The chemical potential along the grain boundary is proportional to the grain-boundary traction~\cite{Chuang:DiffusiveCrack} or $\sigma_{xx}$ and hence the chemical potential along the grain boundary changes abruptly at the dislocation.  Ga atoms in the grain boundary respond by jumping quickly from above the dislocation line to below it.  This, in turn, moves the dislocation down, yet preserves the stress discontinuity.  This explains why the dislocation climbs down at a fixed rate.  How fast does the dislocation climb?  This can be determined by solving the coupled elasticity/diffusion problem.  A similar problem was addressed by Antipov {\it et al.}~\cite{Antipov:IntegroEquation} in the context of diffusive crack growth along a grain boundary subjected to an applied stress~\cite{Chuang:DiffusiveCrack}.  The steady-state dislocation climb velocit
y $V$ in this model can be approximated as 
\begin{equation}
\label{eq:ClimbVelocity}
V \approx  \frac {\Omega D_{gb}} {k T} \frac {E b} {(1-\nu^2) l_{c}^2},  
\end{equation}
where $\Omega$ is the atomic volume of the species with grain boundary diffusivity $D_{gb}$, $kT$ is the thermal energy, $b$ is the Burgers vector and $E$ and $\nu$ are the Young's modulus  and  Poisson's ratio.  $l_{c}$ is a characteristic length associated with the jump in stress across the dislocation and should be of order of the dislocation core size (i.e., a few \AA) and can be found by solving the singular coupled elasticity/diffusion problem~\cite{Antipov:IntegroEquation}.  Using values for Ga in Al in Eq.~(\ref{eq:ClimbVelocity}) yields $V \approx 0.1$ m/s, which is consistent with the dislocation climb velocity in the present simulations.  

It is interesting to note that in the absence of an applied strain, no dislocation forms [Fig.~\ref{fig:LME_asy}(a)] and the Ga penetration rate decreases with time (Fig.~\ref{fig:Prate_asy}).  However, when a strain is applied, dislocations form and climb at fixed rate [Fig.~\ref{fig:LME_asy}(b) and (c)] and the Ga penetration rate is time independent (Fig.~\ref{fig:Prate_asy}).  This suggests that the constant Ga penetration rate observed in the strained solid is associated with the fixed rate of ``climb'' of dislocations.  To examine the relationship between dislocation propagation and Ga penetration, we performed an additional simulation in which we applied a strain (corresponding to 500 MPa) long enough to nucleate a dislocation ($\sim 5$ ns) and then removed the applied strain and continued the simulation.  In this case, no additional dislocations form but the single dislocation continues to ``climb'' down the grain boundary at nearly the same constant rate as when a strain is applied.  However, the Ga penetration depth versus time is sub-linear (Fig.~\ref{fig:Prate_asy}).  We note that the dislocation can propagate down the grain boundary either by Ga atoms or even by Al atoms jumping across the stress discontinuity set up at the dislocation line.  (The latter case is similar to classical diffusive crack growth~\cite{Chuang:DiffusiveCrack}.)  In the case of Fig.~\ref{fig:LME_asy}(d), the dislocation climbs down via Al atom hopping at the dislocation on the grain boundary rather than Ga, since Ga transport can not keep up with the dislocation climb without the aid of residual stress.  When this happens, the Ga penetration rate (which slows in time) and the dislocation climb rate (fixed in time) are decoupled.  Therefore, the applied strain plays two essential roles:  to aid the nucleation of dislocations at the grain boundary and to keep the grain boundary open to allow sufficiently fast Ga transport enough to move with the dislocation.  

In summary, our simulations demonstrate that application of a stress significantly promotes liquid metal penetration along grain boundaries, resulting in a change from a diffusive to fixed rate penetration mode.  This is consistent with experiments that show that stresses accelerate Ga penetration and lead to a constant penetration rate~\cite{Hugo:AlGaTEM, Kozlova:AlGaSEM, Pereiro-Lopez:AlGaPRL2005, Pereiro-Lopez:AlGaSRMR}.  The simulation also confirm the microtomography observations that Ga penetration leads to grain boundary opening~\cite{Pereiro-Lopez:AlGaPRL2005}, and electron microscopy observations of moving stress fields during Ga penetration ~\cite{Hugo:AlGaTEM}.  While consistent with the experimental observations, the present results are not consistent with several theories of LME~\cite{Joseph:LMEreview}.  A new picture of LME emerges.  First, Ga diffuses down the grain boundary in Al below the liquid groove root and causes stresses large enough to nucleate a dislocation in the grain boundary.  The first dislocation ``climbs'' down by stress-enhanced Ga hoping across the dislocation core, leaving a tail of Ga behind.  This Ga hopping leads to a constant dislocation climb rate that is applied stress-independent. Once the dislocation moves far enough from the groove root, another dislocation is nucleated.  It too climbs down the grain boundary at the same rate, resulting in a uniform spacing of climbing dislocations.  With Ga at the grain boundary, applied strains enhance the grain boundary opening and in turn more Ga is inserted from the liquid groove into the grain boundary to relieve the residual stress (i.e., Ga layer thickening process).  The Ga penetration rate mirrors the dislocation climb rate and hence is time independent. In order for LME to occur, the solute must diffuse quickly in the grain boundary, a stress must be applied  to nucleate dislocations and keep the grain boundary open, and the solute must be capable of creating grain boundary decohesion at sufficient concentrations.



\begin{acknowledgments}
The authors gratefully acknowledge the support of the US Department of Energy Grant DE-FG02-011ER54628 and its Computational Materials Science Network.
\end{acknowledgments}



\appendix




\end{document}